\documentclass[prd,nofootinbib,preprintnumbers]{revtex4-2}

\usepackage{amsfonts,amsmath}
\usepackage[colorlinks=true,linkcolor=red,citecolor=blue]{hyperref}
\usepackage[parfill]{parskip}
\usepackage{orcidlink}
\usepackage{subcaption}
\usepackage{booktabs}
\usepackage{graphicx}
\usepackage{float}

\usepackage[T2A]{fontenc}

\newcommand{\be}{\begin{eqnarray}}
\newcommand{\ee}{\end{eqnarray}}
\newcommand{\ba}{\begin{array}}
\newcommand{\ea}{\end{array}}
\newcommand{\tr}{\mbox{\rm tr}}
\newcommand{\nn}{\nonumber}

\newcommand{\bi}{\begin{itemize}}
\newcommand{\ei}{\end{itemize}}

\begin{document}

\preprint{IPARCOS-UCM-26-046}

\title{Leading Logarithms in the Effective Theory of Gravity: corrections to the scalar mass}

\author{
Diana V. Mysliaeva$^{1,2}$\orcidlink{0009-0001-6562-7096},
Kirill M. Semenov-Tian-Shansky$^{3,4}$\orcidlink{0000-0001-8159-0900},
Dmitrii V. Vasilev$^{1,2,4,5}$\orcidlink{0009-0007-2052-1611},
Alexey Vladimirov$^{6}$\orcidlink{0000-0001-5449-194X}
}
\affiliation{
$^1$ Saint-Petersburg State University, St. Petersburg, 199034, Russia \\
$^2$ ITMO University, St.~Petersburg,
 197101, Russia \\
$^3$ Kyungpook National University, Daegu 41566, Korea \\
$^4$ NRC ``Kurchatov Institute''-PNPI, Gatchina 188300, Russia \\
$^5$ Steklov Mathematical Institute, Fontanka 27, St. Petersburg 191023, Russia\\
$^6$ Departamento de Física Teórica \& IPARCOS, Universidad Complutense de Madrid, E-28040 Madrid,
Spain
}

\begin{abstract}

In a low-energy effective field theory (EFT), scaling logarithmic terms arise from the renormalization of higher-dimensional couplings that absorb divergences generated by loop integrals with lower-order vertices. Consequently, leading logarithms (LLs) are generated entirely by the leading-order Lagrangian, rendering them universal low-energy predictions. Adapting techniques developed for chiral perturbation theory, we analyze the series of LLs in the EFT of general relativity. As a concrete application, we compute LL corrections to the scalar pole mass up to three-loop order and demonstrate that they can be determined recursively to arbitrary loop order. To confirm the physical validity of our results, we carry out the calculation using a general gauge-fixing parameter and explicitly verify, up to two loops, its complete cancellation in the physical observable.
\end{abstract}

\date{\today}

\maketitle

\section{Introduction}


Following the effective field theory (EFT) philosophy of S.~Weinberg~\cite{Weinberg:1978kz}, it is now well established \cite{Donoghue:1994dn} that general relativity admits a systematic quantum treatment at low energies despite its perturbative nonrenormalizability. In this approach, the action 
includes an infinite series of terms consistent with diffeomorphism invariance and ordered by their mass dimension. The Einstein--Hilbert gravity action represents the leading term of this series. The higher-order (higher-curvature) operators enter with unknown coupling constants of decreasing mass dimension, which encode short-distance physics and are to be determined from experimental measurements. The quantum-loop effects are also ordered by dimension and induce the running of low-energy constants. Altogether, this enables controlled and predictive quantum calculations for gravitational observables. See Refs.~\cite{Donoghue:1995cz, Burgess:2003jk, Donoghue:2012zc, Donoghue:2017pgk, Burgess:2020tbq, Donoghue:2022eay} for comprehensive reviews.


The necessity to introduce an infinite tower of coupling constants limits the predictive power of any low-energy EFT. However, it still remains far from 
being negligible. 
First of all, the EFT framework allows a systematic calculation of any observable in terms of a few universal parameters. Second, it predicts non-analytic contributions to observables, induced by quantum-loop corrections. While these non-analytic terms can take various forms, the most common are logarithms of ratios of external kinematic parameters, such as particle masses and momenta. Third, the EFT approach guarantees a consistent evolution of low-energy constants and observables under changes in the renormalization scale $\mu$. These corrections also take the form of logarithms and are generated by the ultraviolet asymptotics of loop integrals. In certain cases, such as the running of masses \cite{Bijnens:2014ila} or form factors \cite{Kivel:2009az}, the scaling logarithms are in one-to-one correspondence with logarithms of kinematic variables.

The series of scaling logarithms, $\log\mu$, forms a particularly universal class of quantum contributions in any EFT. The reason is that the $k$-th power of the scaling logarithm at the $n$-th order of energy expansion
(with 
$n=1$ being the leading contribution) is generated by loop corrections that involve low-energy constants of at most $(n-k)$-th order. Therefore, the series of leading logarithms (LLs), which is defined as the series of logarithms of maximum 
($n-1$-th) power at each given ($n$-th) order, depends only on the initial action. The series of next-to-leading logarithms depends on the initial and next-to-leading-order action, and so on. Thus, the series of LLs is insensitive to the ultraviolet (UV) completion of the effective theory and therefore constitutes genuine low-energy predictions. This hierarchy is well known and was first understood in the context of Chiral Perturbation Theory (ChPT) \cite{Weinberg:1978kz, Gasser:1983yg}. In ChPT, the logarithmic series reflects the universal long-distance symmetry-driven dynamics of the Goldstone modes and dominates over naive power counting in appropriate kinematic regimes \cite{Gasser:1983kx}.

In renormalizable quantum field theories, the logarithmic corrections are handled with the help of the renormalization group (RG) framework. For the EFT setting, RG techniques were systematically adapted in Refs.~\cite{Colangelo:1995np, Buchler:2003vw,Bissegger:2006ix}
and, from a slightly different viewpoint in Ref.~\cite{Kazakov:1987jp}. A major breakthrough was achieved in Ref.~\cite{Kivel:2008mf}, which showed that RG invariance permits the determination of LLs to arbitrary loop order through a recursive algorithm. The validity of this construction was illustrated independently in Ref.~\cite{Koschinski:2010mr} with a derivation based purely on the fundamental QFT constraints of unitarity, analyticity, and crossing symmetry. The approach was further developed and expanded in a number of subsequent publications \cite{Bijnens:2009zi, Kivel:2009az, Bijnens:2010xg, Polyakov:2010pt, Bijnens:2012hf, Moiseeva:2012zi, Bijnens:2014ila}. The generic structure of the LL approximation for EFTs, along with the explicit analytic resummation of LL contributions in two-dimensional massless toy effective field theory models, was investigated in Refs.~\cite{Polyakov:2018rdp,Linzen:2018pvj,Linzen:2021fua}.

In the EFT of gravity, analogous logarithmic terms arise in quantum corrections to physical observables and can accumulate information across a large hierarchy of scales. While the associated effects are expected to be extremely small in everyday settings ({\it e.g.}, at solar-system scales), understanding their structure and physical interpretation is essential for determining the predictive content of quantum gravity at low energies. This is particularly compelling because LL corrections encode universal, long-range quantum effects, unaffected by unknown Planck-scale physics. Their resummation can reveal nontrivial scale dependence or hidden symmetries of quantum gravity, and potentially uncover new physical phenomena in extreme but classical-looking settings, such as black hole horizons and cosmic strings \cite{Burgess:2003jk}, neutron star binaries \cite{Wang:2019gry}, or the early universe \cite{Miao:2024shs, Miao:2025gzm}.

Despite spectacular advances in computer algebra systems for diagrammatic calculations in quantum gravity \cite{Latosh:2022ydd}, 
brute-force loop computations remain challenging and tedious. The difficulty stems from the proliferation of tensor indices and structures that must be contracted, leading to rapidly growing computational complexity at higher loop orders. Practically, modern calculations of loop corrections in quantum gravity are limited to one- and two-loop orders \cite{Donoghue:1994dn, Donoghue:2022eay, Bjerrum-Bohr:2022blt}. Therefore, recursive methods based on the RG framework for studying quantum corrections--even in the limited form of logarithmic series--may offer a valuable alternative for exploring the properties of quantum gravity, as well as a useful cross-check for explicit computations.

The goal of this work is to adapt the RG framework and recursive computational methods developed in ChPT to the EFT of gravity, enabling the systematic calculation of LL contributions%
\footnote{An alternative application of RG methods in quantum theory of gravity can be found in Ref.~\cite{Solodukhin:2020vuw}.}. As we demonstrate, the conceptual scheme of the computation remains the same, although one encounters a dramatically increased number of diagrams and algebraic structures in comparison with ChPT. Gauge invariance, which is absent in the ChPT case, does not bring any conceptual complications. As a concrete demonstration, we compute the LL corrections to the scalar mass up to three loops and show that these corrections can be determined recursively to an arbitrary loop order. The computation is performed in the harmonic gauge, but we also verify it with a computation in a general covariant gauge (at the two-loop order) and explicitly demonstrate the cancellation of gauge-dependent terms.

The method developed in this work can be applied to any other observable. A particularly interesting target is the quantum corrections to the Newtonian potential, which involve non-analytic $\propto \log^k(q^2)$ terms in momentum space. These terms generate $\propto 1/r^3$ and other long-distance corrections to the potential in position space, and thus are responsible for the leading deviation from the classical part of the Newtonian potential. The direct diagrammatic evaluation of this effect proved to be highly nontrivial, unfolding over a couple of decades (see, {\it e.g.}, \cite{Donoghue:1993eb, Hamber:1995cq, Akhundov:1996jd, Khriplovich:2002bt}), with a fully consistent result finally emerging in Ref.~\cite{Bjerrum-Bohr:2002gqz}%
\footnote{There is an ongoing discussion in the literature concerning which diagrams must be included to ensure the gauge independence of the potential \cite{Shapiro:2008ss}. For a recent calculation of quantum corrections to the Newtonian potential using functional methods, see Ref.~\cite{dePaulaNetto:2021axj}. Related discussions can be found in Refs.~\cite{Akhundov:2006gh,Faller:2007sy}, while Ref.~\cite{Brandt:2022und} presents a calculation in a general background gauge.}.
We regard the present work as a preparatory step toward the computation of the LL series for Newton's law. To this end, we also provide an independent calculation of the one-loop correction to the Newtonian potential, finding complete agreement with Ref.~\cite{Bjerrum-Bohr:2002gqz}. This calculation provides an additional cross-check of our framework.

The discussion is organized as follows. In Sec.~\ref{II}, we briefly review the effective field theory description of gravity coupled to a massive scalar field, introduce the expansion in principal orders, and specify the gauge-fixing and ghost sectors of the theory. In Sec.~\ref{Sec_renorm}, we discuss the power-counting and renormalization structure of the theory and illustrate it 
by computing the one-loop graviton polarization operator.  In Sec.~\ref{Sec_LL_scalar_mass}, we apply the recursive renormalization-group approach to the scalar two-point function and determine the leading-logarithmic corrections to the scalar pole mass, including a check of their gauge-parameter independence. Finally, Sec.~\ref{Sec_Conclusions} summarizes our results and discusses possible further applications of the method.

\section{Gravity as effective field theory} 
\label{II}

In this section we review the construction and main elements of the EFT of general relativity. This material is standard and is presented here for the completeness of exposition. 
The sign conventions and definitions adopted here are identical to those employed in Ref.~\cite{Donoghue:1995cz}.

We work in the spacetime dimension 
$d=4$ 
with the flat metric 
$\eta_{\mu\nu}=\text{diag}(1,-1,-1,-1)$ 
and make use of the system of units in which 
$c=1$ 
and 
$\hbar=1$.
We consider an EFT containing a massless spin-2 field $h_{\mu\nu}$ 
and a massive scalar field $\phi$, assuming that this theory possesses the gauge symmetry under infinitesimal diffeomorphisms
\be
\delta^\epsilon x_\alpha = x^{\epsilon}_\alpha- x_\alpha = \epsilon_\alpha(x).
\label{Def_local_transform}
\ee 
The scalar and tensor fields transform under (\ref{Def_local_transform}) 
according to 
\be
&\delta^{\epsilon}\phi(x)=\partial_{\alpha}\phi(x)\epsilon^{\alpha}(x),\\
&\delta^{\epsilon}h_{\nu\mu}(x) =\left[\eta_{\mu \alpha}+ h_{\mu \alpha}(x)\right]\partial_{\nu}\epsilon^{\alpha}(x) +\left[\eta_{\nu \alpha}+ h_{\nu \alpha}(x)\right]\partial_{\mu}\epsilon^{\alpha}(x)+ \epsilon^{\alpha}(x)\partial_{\alpha}h_{\nu \mu}(x), \label{h_transform}
\ee
which correspond to an infinitesimal gauge transformation with parameter 
$\epsilon^{\alpha}(x)$. 
Note that the transformation law of an arbitrary rank-$2$ tensor in the infinitesimal form is given by 
\be
\delta^{\epsilon} A_{\mu \nu}= A_{\mu \alpha}\partial_{\nu}\epsilon^{\alpha}(x) +A_{\alpha\nu}\partial_{\mu}\epsilon^{\alpha}(x)+\epsilon^{\alpha}(x)\partial_{\alpha}A_{\mu \nu}(x),
\ee 
which differs from 
Eq.~\eqref{h_transform}. 
The additional terms arise because the tensor field $h_{\mu\nu}$ 
is regarded as a fluctuation of the spacetime metric $g_{\mu\nu}$ 
around the flat background,
\(
g_{\mu\nu}=\eta_{\mu\nu}+h_{\mu\nu},
\)
while the background metric 
$\eta_{\mu\nu}$ 
is kept invariant under gauge transformations. Therefore, we define the transformation of the field $h_{\mu\nu}$ 
as 
\be
\delta^{\epsilon}h_{\mu\nu}=\delta^{\epsilon}(g_{\mu\nu}-\eta_{\mu\nu})=\delta^{\epsilon}g_{\mu\nu}, 
\ee
see e.g. Sec. 22 of 
\cite{Schwartz2013QFT} 
for a more detailed discussion.

The EFT point of view encourages us to consider the most general set of interactions consistent with the underlying symmetries. Let us 
present the 
form of the action. Contributions to the pure gravity sector can be collected step-by-step into the so-called ``principal'' orders in powers of momenta, 
$O(p^{2n})$, $n=1,\,2,\, \ldots\,$, 
in close analogy with the chiral expansion in ChPT \cite{Weinberg:1978kz,Gasser:1983yg}. 
For convenience, in what follows we rescale the tensor field as 
$h_{\mu\nu}\to \kappa h_{\mu\nu}$, 
where 
$\kappa$ 
is a constant of dimension $[\kappa]=-1$ that is 
specified in \eqref{kappa_plank}.
Suppressing indices, one can schematically write 
the gravitational EFT action as a series:
\be 
   S_{\text{grav}} = \int d^{4}x\Bigg[\overbrace{\left(h \partial^2h+ \frac{c^{(1)}_{1}}{M} h^2\partial^2h+ \frac{c^{(1)}_{2}}{M^2} h^3\partial^2h +\ldots\right)}^{\mathcal{L}_{\text{grav}}^{(1)}}+\overbrace{\left(\frac{c_{0}^{(2)}}{M^2}h \partial^4h+ \frac{c^{(2)}_{1}}{M^3} h^2\partial^4h+ \frac{c^{(2)}_{2}}{M^4} h^3\partial^4h +\ldots\right)}^{\mathcal{L}_{\text{grav}}^{(2)}}+\ldots\Bigg],
   \label{Def_Sgrav}
\ee 
where 
$M \sim M_{\text{Planck}}$ 
is the scale of the UV 
completion of the theory and $\mathcal{L}^{(n)}_{\text{grav}}$ 
denotes the Lagrangian of $n$-th 
principal order that contains  $2n$ derivatives. 
We obtain the set of dimensionless constants 
$\{c^{(n)}_{k j}\}_{\text{grav}}$, 
each corresponding to a possible Lorentz-invariant operator 
built of 
$k+2$ 
fields and involving 
$2n$ 
derivatives with index 
$j$ 
distinguishing operator species. The corresponding set of operators can be reduced to some minimal basis by performing integrations by parts and employing suitable algebraic identities. However, for our purposes, it is sufficient to work with an overcomplete operator basis.



The pure matter sector 
\(S_{\text{matter}}(\phi)\) 
is expanded similarly:
\begin{align}
S_{\text{matter}}\,=& \int d^4x \Bigg[\overbrace{\left(\frac{1}{2}\eta^{\mu \nu}\partial_{\mu}\phi\partial_{\nu}\phi-\frac{m^2}{2} \phi^2\right)}^{\mathcal{L}^{(1)}_{\text{matter}}} \nonumber \\
+\, & \overbrace{\left( \phi\left[\frac{c_{01}^{(2)}}{M^2}m^4+\frac{c_{02}^{(2)}}{M^2}\partial^4+\frac{c_{03}^{(2)}}{M^2}m^2 \partial^2\right] \phi+ \phi^2\left[\frac{c_{11}^{(2)}}{M^3}m^4+\frac{c_{12}^{(2)}}{M^3}\partial^4+\frac{c_{13}^{(2)}}{M^3}m^2 \partial^2\right] \phi +\ldots \right)}^{\mathcal{L}^{(2)}_{\text{matter}}}+\ldots\Bigg],
\label{Def_Smatt}
\end{align}
bringing the set of parameters 
$\{c^{(n)}_{k j}\}_{\text{matter}}$ 
corresponding to   Lorentz-invariant operators 
constructed from 
$k+2$ 
scalar fields and involving 
$2n$ 
derivatives
(and/or powers of the mass  
$m$)  
with index 
$j$ 
labeling the operator species. 
We assume that no self-interactions appear at the 
$n=1$ principal order
in 
\(S_{\text{matter}}\).

The interaction sector 
\(S_{\text{int}}(\phi, h_{\mu\nu})\) 
admits a similar expansion, characterized by its own set of coupling constants. The gauge symmetry of the theory 
\eqref{h_transform}
imposes a restriction on the possible form of the Lagrangian, which leads to some couplings being dependent. In this work, we 
consider the minimal coupling of gravity to matter. We note, however, that the particular form of 
$S_{\mathrm{grav}}^{(1)}$ 
implicitly requires the absence of bulk terms linear in 
$h_{\mu\nu}$, 
which is equivalent to the flat metric being a stationary point of the gravitational action
\(
\left.
\frac{\delta S_{\mathrm{grav}}^{(1)}}{\delta h_{\mu\nu}}
\right|_{h_{\mu\nu}=0}=0.
\) Therefore, to consistently expand around the flat space, throughout this work we omit a possible zero-derivative contribution to the gravitational part\footnote{
If such a contribution was included, the diffeomorphism invariance would restrict it to the cosmological-constant term,
\(
S_{\mathrm{grav}}^{(0)}
=
\Lambda\int d^4x\,\sqrt{-g}.
\)
Such a term generates contributions linear in $h_{\mu\nu}$, 
reflecting the fact that the flat metric is no longer a solution of the vacuum Einstein equations. 
}. Our leading action reads: 
\be
    S^{(1)}_{\rm grav \, + \,matter}=\int d^4x\sqrt{-g}\left\{\frac{2}{\kappa^2}   \, R +\frac{1}{2} g^{\mu \nu}\partial_{\mu}\phi \partial_{\nu}\phi-\frac{1}{2}m^2\phi^2\right\}. 
\label{S1_grav_plus_matter}    
\ee
Here 
$g_{\alpha \mu}g^{\mu\beta} = \delta^{\beta}_{\alpha}$, 
$g = \text{det}(g_{\mu\nu})$ 
and 
$R$ 
is the Ricci scalar constructed from the Levi--Civita connection 
$\Gamma^\rho_{\alpha \beta}$ 
depending on 
$g_{\mu\nu}$. 
Our conventions for the Riemann tensor, the Ricci tensor and the scalar curvature are the following: 
\be
R^\rho_{\ \sigma\mu\nu} = \partial_\mu \Gamma^\rho_{\nu\sigma} - \partial_\nu \Gamma^\rho_{\mu\sigma} + \Gamma^\rho_{\mu\lambda} \Gamma^\lambda_{\nu\sigma} - \Gamma^\rho_{\nu\lambda} \Gamma^\lambda_{\mu\sigma},\quad  R_{\mu\nu} = R^\alpha_{\ \mu\alpha\nu}, \quad R = g^{\mu\nu} R_{\mu\nu}.
\ee
All these quantities must be considered as series in the field 
$h_{\mu \nu}$, 
see Appendix~\ref{append:vertices}. 
The expansion of 
\eqref{S1_grav_plus_matter} 
relates all couplings 
$\{c^{(1)}_{k j}\}$ 
to a single non-trivial 
coupling constant 
$\kappa$ 
that can be fixed by matching with the  tree-level Newton potential \cite{Donoghue:1994dn}:
\be
   \kappa^2 = 32 \pi G \equiv 32 \pi M_{\rm Planck}^{-2}. \label{kappa_plank}
\ee
One can carry out this procedure for higher orders as well. For example, it is well known that, at the next principal order, the gravitational sector in 
$d=4$ 
takes the form
\be 
    S_{\text{grav}}^{(2)}= 
    \int d^4x\sqrt{-g}\left\{a^{(2)} R^2+ \Tilde{a}^{(2)} R_{\mu\nu} R_{\alpha \beta} g^{\alpha \mu } g^{\beta \nu} \right\}.
\ee
Generally speaking, the gauge invariance imposes relations on the coupling constants, which could be used to reduce their number and simplify the action. 
However, the recurrence procedure described below does not require accounting these relations; invoking them would, in fact, only complicate the subsequent analysis.
Therefore, we consider the higher-principal-order actions with a redundant set of couplings 
\(\{c^{(n)}_{kj}\}\) 
without taking into account possible relations among them.


Let us briefly discuss the quantization of the theory. Gauge invariance implies that the corresponding path integral must be restricted to integration over gauge orbits only. We implement the standard Faddeev-Popov procedure and fix the gauge using the so-called ``harmonic'' gauge-fixing function 
\be 
    G^{\alpha} = \partial^{\mu}h^{\alpha}_{\mu}-\frac{1}{2}\partial^{\alpha}h^{\lambda}_{\lambda}.
\ee
The action 
(\ref{S1_grav_plus_matter}) is 
modified by adding the gauge fixing and ghost terms:
\be
S^{(1)}=S^{(1)}_{\text{grav+matter}}+S^{(1)}_{\text{gf}}+S^{(1)}_{\text{ghost}},
\label{Complete_S_1}
\ee
where
\be
    S^{(1)}_{\text{gf}}= && \frac{1}{\xi}\int d^4x\left\{G^{\mu}G_{\mu} \right\};
\label{S1_gf}    \\
    S^{(1)}_{\text{ghost}} = && \int d^4x d^4y \left\{
    \Bar{c}^{\mu}(x)  \,  \frac{\delta G^{\nu}
    (h_{\alpha \beta}+\delta^{\epsilon} h_{\alpha \beta}(x))}{\delta\epsilon^{\mu}(y)}c_{\nu}(y)\right\} \nonumber \\
    = &&\int d^4x\left\{\Bar{c}^{\nu}\partial^2c_{\nu}+\kappa\Bar{c}_{\nu}\left[ \partial_{\alpha}\left( G^{\nu}\right)+h^{\nu}_{\alpha} \partial^2+ G_{\alpha}\partial^{\nu}+\left(\partial^{\sigma}(h^{\nu}_{\alpha})-\partial^{\nu}(h^{\sigma}_{\alpha})+\partial_{\alpha}h^{\nu \sigma}\right)\partial_{\sigma} \right]c^{\alpha}\right\}.
\label{S_Ghost_sector}    
\ee
The quadratic part of the 
complete  
$n=1$ 
action  
(\ref{Complete_S_1})
reads: 
\be
    S^{(1)}_{\text{quad}}= \int d^4x\left\{ \frac{(\partial_{\alpha}h^{\nu\mu})^2}{2} +\left(1-2\xi\right)\frac{(\partial_{\mu} h^{\lambda}_{\lambda})^2}{2\xi} +(1-\xi)\frac{(\partial_{\mu}h^{\mu \alpha})^2-\partial_{\alpha}h^{\lambda}_{\lambda}\partial_{\mu}h^{\mu \alpha}}{\xi}+\frac{(\partial_{\mu}\phi)^2}{2}-\frac{m^2\phi^2}{2}+\Bar{c}^{\nu}\partial^2c_{\nu}\right\}.
    \label{S1_quad}
\ee
This allows to express the corresponding propagators.
For the graviton propagator, in arbitrary 
\(  d\) 
dimensions, we recover the result of Ref.~\cite{Brandt:2022und}:
 \be
 &&
\mathcal{D}_{\alpha \beta, \, \gamma \delta}(q)=
\frac{i}{q^2+i 0}
\left[\frac{1}{2}\left(\eta_{\gamma \beta} \eta_{\delta \alpha}+\eta_{\gamma \alpha} \eta_{\delta \beta}-\frac{2 \eta_{\gamma \delta} \eta_{\alpha \beta}}{d-2}\right)+\frac{\xi-1}{2 (q^2+i0)}\left(q_\delta q_\beta \eta_{\gamma \alpha}+q_\delta q_\alpha \eta_{\gamma \beta}+q_\gamma q_\beta \eta_{\delta \alpha}+q_\gamma q_\alpha \eta_{\delta \beta}\right)\right].
\nn \\ &&
\ee

The ghost and scalar propagators take the standard form
\be 
    \mathcal{D}_{\alpha\beta}(q)=\frac{i\eta_{\alpha \beta}}{q^2+i0 }, \quad \mathcal{D}(q)=\frac{i}
    {q^2-m^2+i0    }.
\ee
The complete expression for 
$S^{(1)}$ 
contains interaction vertices with arbitrarily many fields 
$\phi$, $c_\mu$, 
and 
$h_{\mu\nu}$. 
We provide explicit expressions for some of the interaction vertices from 
$S^{(1)}$ in  
Appendix~\ref{append:vertices}.

It is worth noting that the formulation of gravity as an EFT presented here differs in several technical aspects from the standard background-field approach of 
Ref.~\cite{Donoghue:1994dn}. The background approach implies that the gravitational field is expanded around the background metric 
$\bar g_{\mu\nu}$,  
which is subsequently decomposed around 
flat space-time as 
$\bar g_{\mu\nu} = \eta_{\mu\nu} + \kappa H_{\mu\nu}^{\rm ext}$.  
Then, by computing the 1PI diagrams, one reconstructs the effective action of gravity. It is however overwhelming for our task, because in the effective action the renormalization terms are multiplied together to form a complete renormalized action, and thus an additional algebra is required to extract the renormalization of each individual element.  Meanwhile, we are seeking only for the renormalization counter terms for coupling constants, which allow us to deduce the LL coefficients, as described below. Therefore, it is sufficient to consider only the pure quantum part 
taking corresponding UV singular parts of the loops. 



As a consistency check of our conventions and Feynman rules, in Appendix~\ref{App_one_loop_diags}, 
we reproduce the well-known one-loop quantum correction to the Newtonian potential 
\cite{Bjerrum-Bohr:2002gqz}.

\section{ Renormalization and power counting}
\label{Sec_renorm}

The organization of the action by principal order provides the operational principle of the theory. It enables a systematic separation of contributions based on their relative importance at a given energy scale. Consequently, the dynamics can be formulated through a hierarchy of operators, with higher-order terms encoding progressively suppressed effects. In this section, we briefly revisit the this concept and demonstrate how it helps to extract higher dimensional contributions with minimal efforts.

In general, all kinematic variables and the  masses 
$m$ 
are considered to be of the order of the low-energy scale $E$. Therefore, the computation of a Green's function or an amplitude $\mathcal{M}$ 
with effective action 
(\ref{Def_Sgrav}), 
(\ref{Def_Smatt}) 
produces a series of terms with increasing powers of $E$, compensated by inverse powers of 
$M$ ($\sim M_{\text{Planck}}$). 
In the absence of the dimension-$0$ operators, only a finite number of vertices 
($\mathcal{L}^{(n)}$) 
contribute to an observable at each order of expansion. 


Consider a momentum-space amplitude 
$\mathcal{M}_K$ 
with 
$K$ 
external legs. Its mass dimension is 
$[\mathcal{M}_K]=4-K$.
At the tree-order, the leading order in the energy expansion is generated by the corresponding operator in 
$S^{(1)}$, 
while further contributions from 
$S^{(n)}$ are suppressed. 
Factoring out the overall mass dimension, we can write
\be
\mathcal{M}_K \sim M^{4-K}\left(E^2 / M^2+E^4 /M^4+\ldots\right),
\ee
so that the remaining expansion is organized in powers of 
$E/M$, 
and it suffices to track the powers of 
$E$. 

Consider a generic diagram contributing to $\mathcal{M}_K $  
with 
$L$ loops and 
$I$ 
propagators, and let 
$N_k$ 
denote the number of vertices with overall $k$-th power of energy scale.

By using the 
topological identity
\be
L = I - \sum\limits_{k=2,4,\ldots} N_k + 1,
\ee
one may conclude that the contribution of such 
diagram into $\mathcal{M}_K $ scales as
%
\begin{align}
    M^{4-K} \left(E / M \right)^{W}, \quad \text{with} \ \ \ W= 2L +2 +\sum_{k=2,4,\ldots}N_{k}(k-2). \label{power_couting}
\end{align}
Thus, the UV divergent part of this diagram contributes to the renormalization of 
$S^{(W/2)}$ 
action. 

An important point to be taken into account is that in the dimensional regularization, that we utilize to regularize UV divergences with $d=4-2\varepsilon$, the mass dimension of coupling constants is modified by scaling factor preserving its canonical dimension. Therefore,
\be
c^{(n)}_k(\mu) \to c^{(n)}_k(\mu) \,\mu^{k\varepsilon}. \label{scaling_of_c}
\ee
We stress that, according to this prescription, each diagram with $K$ external legs and $L$ loops carries an overall factor
$\mu^{(K+2L-2)\varepsilon}$.

After taking the limit 
$\varepsilon \to 0$, 
UV divergences arise as poles, while the polynomial dependence on the energy variable 
$E$ is completely determined by the powers 
$E^{W}$, 
and we proceed to compute 
$\mathcal{M}$ 
order by order. The lowest contribution 
$E^2$ 
arises for 
$L=0$ and $k=2$, 
and corresponds to the tree-level diagram computed using 
$S^{(1)}$. 
The next order 
$E^{4}$ 
arises either from 
$L=1$ 
diagrams with 
$N_{k>2}=0$, 
or from tree-level diagrams 
($L=0$) 
with four-derivative vertices, 
$N_{4}=1,N_{k \neq 4}=0$. 
This means that at this order, we must include all tree-level vertices from 
$S^{(2)}$ 
and construct all possible one-loop graphs using vertices from $S^{(1)}$. 
At an arbitrary order 
$E^{2n}$, 
one must include the tree-level contribution coming from 
$S^{(n)}$ 
and construct all loop diagrams up to 
$L = n-1$ 
using vertices from previous principal orders. A fixed number of external lines and loops in a diagram limits the number of vertices, and therefore only a finite number of operators from $S$ contribute to the computation.

Moreover, this grouping procedure facilitates the renormalization of the theory. At second order $E^4$, the one-loop contribution contains a pole divergence of the form 
$b^{(2)}_{K-2}[c^{(1)}_{k}] / \varepsilon$ with $b[c^{(1)}_{k}]$ being a polynomial function of the couplings 
$c^{(1)}_{k}$. 
As we have seen, it must be accompanied by the 
$ c^{(2)}_{K-2}$ 
term and the divergence can be eliminated via the shift of the coupling 
\be
c^{(2)}_{K-2} \to c^{(2)}_{K-2} - b^{(2)}_{K-2}[c^{(1)}_{k}] / \varepsilon
\ee
in the MS  scheme. In the same way, we can eliminate the overall divergence in any Green's function at a given principal order. Of course, this argument does not clarify the cancellation of non-local subgraph divergences, but this in fact proceeds in exactly the same manner as in renormalizable theories, see {\it e.g.} Ref.~\cite{Kazakov:2008tr}. 

As an illustrative example, we 
present
here the one-loop renormalization of the 1PI two-graviton function in the pure gauge sector. The second principal order contains two diagrams: a one-loop diagram constructed from the first-principal-order vertices and a tree-level diagram containing the second-principal-order couplings.    
\begin{figure}[!ht]
\begin{center}
\includegraphics[width=0.3\textwidth]{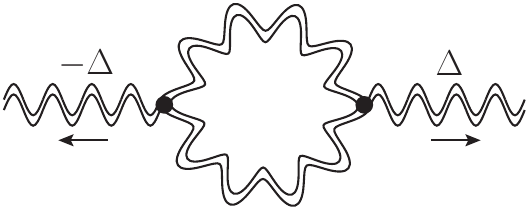} \ \ \ \ \ 
\includegraphics[width=0.3\textwidth]{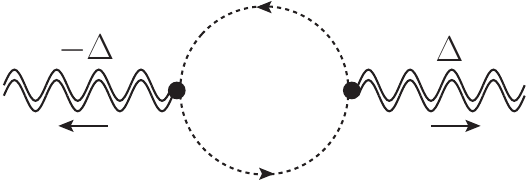}
\end{center}
\caption{One-loop contributions to the 1PI graviton polarization operator with vertices taken from $S^{(1)}$. Tadpole diagrams, which vanish in dimensional regularization, are not shown. }
\label{Fig_pol}
\end{figure}

The divergent part of the one-loop 1PI graviton polarization operator computed in a general covariant gauge admits the following decomposition in the basis of the Nieuwenhuizen projection operators (see Appendix~\ref{App_Nieu}): 
\be
&&
 \Pi^{\rm div}_{\mu \nu, \, \alpha \beta}(\Delta)= 
 \frac{\kappa^2}{
 16 
 \pi^2\varepsilon}
 \Bigg[ 
 \left( \frac{61}{120} -\frac{1}{2}(1-\xi)+\frac{1}{6}(1-\xi)^2  \right) \left(\Delta^2\right)^2 P^2_{\mu \nu, \, \alpha \beta}(\Delta) \nn \\ &&
 + \left( \frac{13}{12} -2(1-\xi)+ \frac{5}{3}(1-\xi)^2 \right)
 \left(\Delta^2\right)^2 P^0_{\mu \nu, \, \alpha \beta}(\Delta)
 + \frac{1}{6}(1-\xi) \left(\Delta^2\right)^2 \bar{\bar{P}}^0_{\mu \nu, \, \alpha \beta}(\Delta)
 \Bigg].
 \label{GravPol_1PI}
\ee
The expansion coefficients in (\ref{GravPol_1PI}) are explicitly
$\xi$-dependent, as expected for an off-shell 1PI two-point function. 
Moreover, besides the transverse spin-2 sector proportional to the $P^2$ and scalar components proportional to the $P^0$, the polarization operator contains a longitudinal mixing contribution proportional to the
$\bar{\bar{P}}^0$
projector. This makes our graviton polarization operator non-transverse for
$\xi \ne 1$%
\footnote{In the background field method the effective action is invariant under background gauge transformations, which enforces transversality of the background two-point function and excludes longitudinal mixing structures in the polarization operator; see \cite{Brandt:2022und}. In the conventional formulation, by contrast, the 1PI self-energy is an off-shell quantity that is not constrained to be transverse.}. Contraction with the physical polarizations $\epsilon_{\mu\nu}^{(\pm)}$ selects the coefficient in front of the projector $P^2$ and vanishes on the mass shell $\Delta^2=0$, reproducing the well-known result of \cite{tHooft:1974toh}.
The $\sim \bar{\bar{P}}^0$ term in (\ref{GravPol_1PI}) is not a physical observable by itself. 
Within the BRST framework \cite{Barnich:2000zw}, such longitudinal structures belong to the gauge-fixing (BRST-exact) sector of the theory and do not correspond to propagating degrees of freedom. Physical information is extracted from the on-shell scattering amplitude, where longitudinal components cancel in accordance with the Slavnov-Taylor identities \cite{Kazakov:2000mu}.  Using \eqref{GravPol_1PI}, we can determine the counter term coming from $S^{(2)}_{\text{grav}}$:
%
\be
&& \int d^{4}x\left\{\frac{1}{2} h^{\mu \nu} \Pi^{\text{div}}_{\mu \nu,\alpha \beta}(-i\partial)h^{\alpha \beta} \right\} \subset \nonumber \\
&& \int d^4x\sqrt{-g}\left\{a^{(2)} R^2+\Tilde{a}^{(2)} R_{\mu\nu} R_{\alpha \beta} g^{\alpha \mu } g^{\beta \nu}\right\}+\int d^4x\left\{\kappa^2\Bar{a}^{(2)}h^{\mu \nu}\left(P^{0}_{\mu \nu, \alpha \beta}(-i\partial)+\frac{1}{3}\Bar{\Bar{P}}^0_{\mu \nu, \alpha \beta}(-i\partial) \right)\partial^4h^{\alpha \beta} \right\}.
\label{Div_Grav_self_energy}
\ee
The divergences of (\ref{Div_Grav_self_energy}) are absorbed by renormalization of the couplings
\be
a^{(2)} \to a^{(2)}- \frac{(20\xi^2-30\xi-9)}{
120}\frac{1}{16\pi^2 \varepsilon}, \quad  \Tilde{a}^{(2)} \to \Tilde{a}^{(2)}- \frac{(20\xi^2+20\xi+21)}{60 }\frac{1}{16\pi^2 \varepsilon}, \quad \Bar{a}^{(2)} \to \Bar{a}^{(2)}-\frac{1-\xi}{4}\frac{1}{16 \pi^2 \varepsilon}.
\ee
We observe that the scalar and the longitudinal mixing sectors contribute to the renormalization of $S^{(2)}_{\text{gf}}$, and the manifest diffeomorphism invariance of the action is lost when $\xi \neq 1$. For this reason, it is more convenient to reconstruct the action ``on the fly,'' proceeding order by order by extracting the divergent parts of the Green functions and assigning to each structure its own coupling $c^{(n)}_{kj}$ together with the corresponding $b$-coefficient multiplying the pole. It may happen that some parts of the action are not reconstructed in this way, since the corresponding operators are not renormalized at the order under consideration. In this case, the important point for our purposes is that the corresponding $b$-coefficients vanish, $b=0$.

\section{Scalar mass running in the  leading logarithm approximation}
\label{Sec_LL_scalar_mass}

In this Section, employing the recursive LL algorithm  summarized in Appendix~\ref{App_rec_sheme}, we compute the LL contribution to the running of the scalar particle mass up to three loops. In EFTs, a running mass refers to a scale-dependent renormalized mass parameter rather than the physical pole mass. Changing the renormalization scale $\mu$ induces a reshuffling of contributions between loop diagrams and local counterterms. The scale dependence of these counterterms compensates that of the loops, ensuring physical observables remain unchanged. In ChPT, for instance, the scale dependence of the chiral logarithms in the pion self-energy is exactly canceled by the running of the corresponding low-energy constants \cite{Gasser:1983yg}. We emphasize that this calculation primarily serves to demonstrate the application of the method using a simple example before it is applied to  physically more relevant observables.

Our analysis closely follows that of 
Sec.~5 of \cite{Bijnens:2010xg}; see also Sec.~3.1 of
\cite{Bijnens:2012hf}. Before applying the recursive LL algorithm, we first illustrate how 
leading logarithms arise from the cancellation of nonlocal divergences at the first nontrivial principal orders.

Let us denote the sum of one-particle-irreducible  diagrams by \( i \Sigma(p^2, m^2) \).
The second principal order contains two contributions: one coming from 
the tree-level and the other from $1$-loop diagrams.
\begin{figure}[H]
\begin{center}
\includegraphics[width=0.35\textwidth]{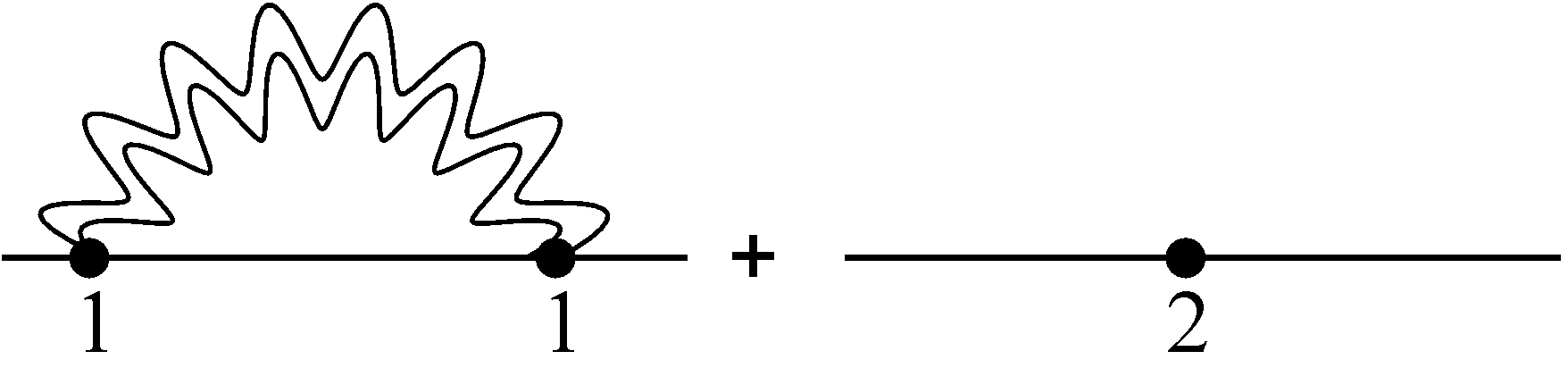} 
\caption{Diagrams contributing to the running of the scalar mass at the second principal order. 
The numbers assigned to the vertices indicate their corresponding principal order.}
\label{Fig_sigma2}
\end{center}
\end{figure}
We focus on the part of the answer that may depend on \( \mu \) and \( \varepsilon \):
\be
\Sigma^{(2)}(p^2,m^2)  && = \    \kappa^2\Bigg[ \underbrace{\left(\frac{1}{\varepsilon}+\log\mu^2 \right) \left(b^{(2)}_{0 1} m^4 + b_{0 2}^{(2)} p^4+b_{0 3}^{(2)} p^2 m^2 \right)}_{1 \ \text{loop}} \nonumber \\
+ && 
\underbrace{ m^4 \left( -  \frac{b^{(2)}_{0 1}}{\varepsilon}+c^{(2)}_{01}(\mu)\right)+ p^4 \left( - \frac{ b^{(2)}_{0 2}}{\varepsilon}+c^{(2)}_{02}(\mu)\right)+ p^2 m^2  \left( -  \frac{b^{(2)}_{0 3}}{\varepsilon} +c^{(2)}_{03}(\mu)\right)}_{\text{tree}}\Bigg]+\ldots .
\label{Sigma2}
\ee
The straightforward calculation yields 
\be
b^{(2)}_{01}= -\frac{1}{16 \pi^2}; \ \ \ b^{(2)}_{02}= 0; \ \ \ b^{(2)}_{03}= \frac{1}{16 \pi^2}.
\ee

The factor \(  \left(1/\varepsilon + \log\mu^2 \right)\) in (\ref{Sigma2}) arises 
from the expansion of \( \mu^{2\varepsilon} / \varepsilon \). 

The pole in (\ref{Sigma2}) cancels due to the renormalization of the constants \( c^{(2)}_{0j}(\mu) \).
Imposing the condition of 
\( \mu \)-independence for each independent Lorentz structure, we obtain
\be
c^{(2)}_{0 j}(\mu)= - b^{(2)}_{0 j} \log(\mu^2). 
\label{rg_1}
\ee
This is the EFT analog of the one-loop running of the coupling constant and the first equation in the system of equations of the generalized renormalization group \eqref{beta_definition_RG}. 

\begin{figure}[H]
\begin{center}
\includegraphics[width=1\textwidth]{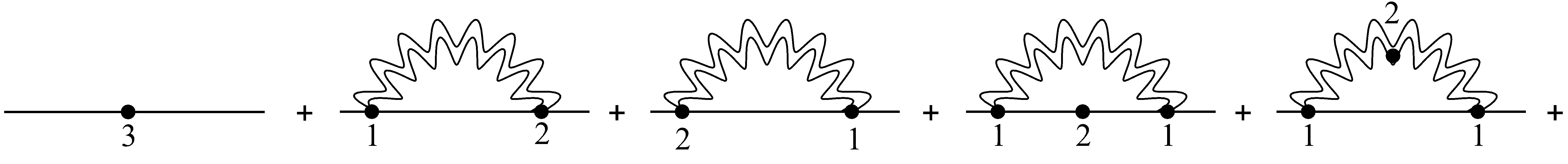} \\
\includegraphics[width=0.6\textwidth]{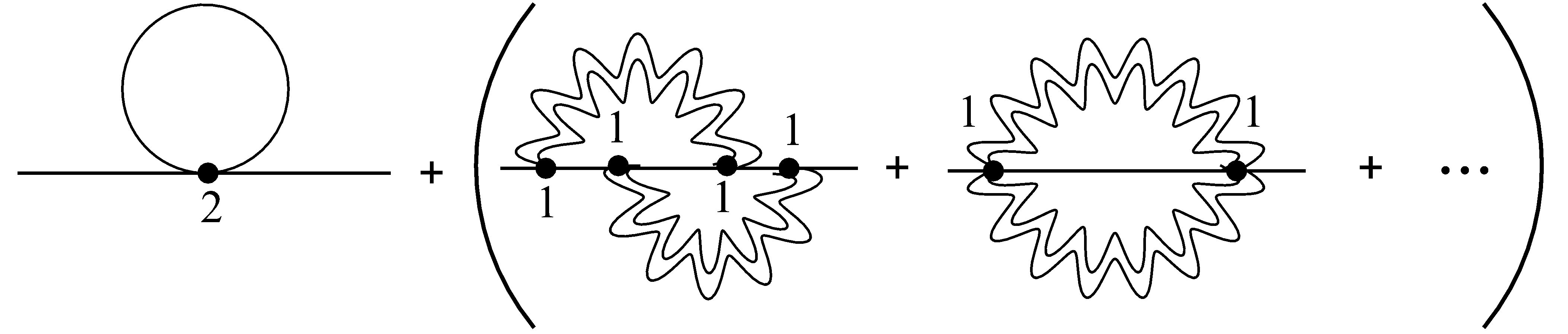} \\
\caption{Tree-level, 1-loop and 2-loop diagrams contributing to the running of the scalar mass at the third principal order. See Eq.~(\ref{sigma_i3}) for the general structure of corresponding contributions. 
}
\label{Fig_sigma3}
\end{center}
\end{figure}

Now we turn to the next order. As explained in Sec.~\ref{Sec_renorm}, in addition to the tree-level and one-loop contributions, we must take into account all two-loop diagrams, see Fig.~\ref{Fig_sigma3}. 

\begin{figure}[H]
\begin{center}
\includegraphics[width=0.8\textwidth]{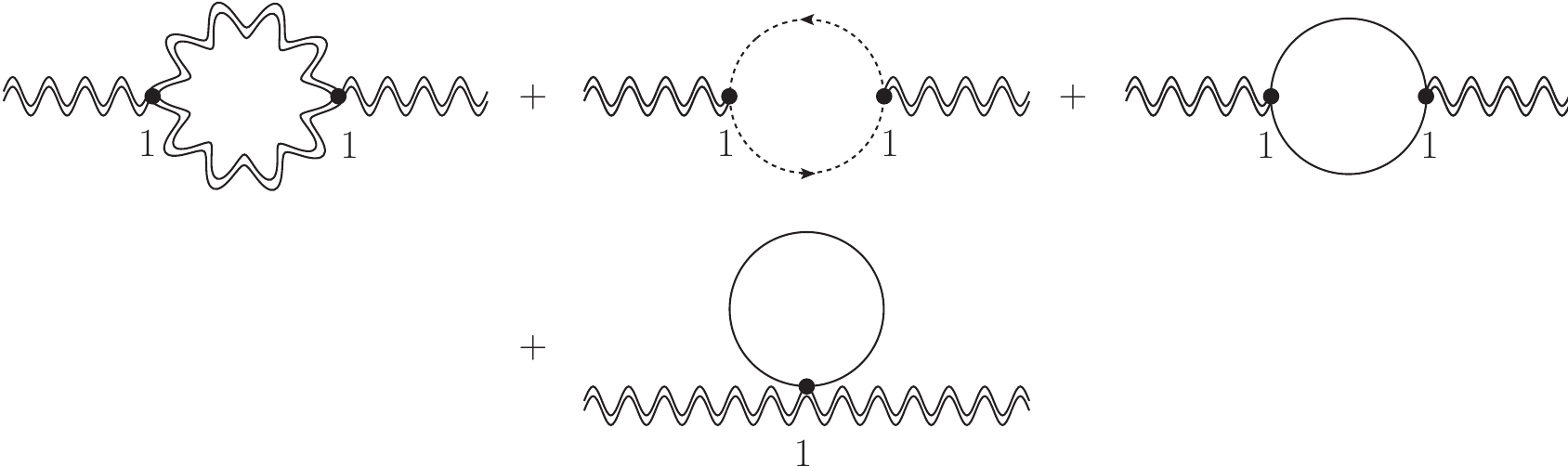 } 
\caption{One-loop diagrams for the second-order gravitational two-point function.}
\label{Fig_2point_principal_order_2}
\end{center}
\end{figure}

At this stage, we need to renormalize all the second-order vertices in Fig.~\ref{Fig_sigma3}. For example, for the second principal order gravitational two-point function in the last graph of the first line of Fig.~\ref{Fig_sigma3}, in addition to the diagrams in Fig.~\ref{Fig_pol}, we compute the divergences of two extra loop diagrams involving scalars (see Fig.~\ref{Fig_2point_principal_order_2}).

Defining the expansion coefficients with respect to the Lorentz structures
\be
\Sigma^{(3)}(p^2,m^2)= \kappa^4 \left[p^6 \sigma^{(3)}_{1} + p^4 m^2 \sigma^{(3)}_{2}+p^2 m^4 \sigma^{(3)}_{3}+m^6\sigma^{(3)}_{4} \right],
\ee
we present the most general form for each of them.
\be
&&\sigma^{(3)}_{i} =  \underbrace{\mu^{4 \varepsilon}\left(\frac{X}{\varepsilon^2}+\frac{Y}{\varepsilon}\right)}_{2 \  \text{loop}}+\underbrace{\sum_{k,j} \alpha_{kj}\frac{\mu^{2 \varepsilon}}{\varepsilon}\bigg(c^{(2)}_{k j}(\mu)-\frac{b_{kj}^{(2)}}{\varepsilon}\bigg)}_{1 \ \text{loop}}+\underbrace{c^{(3)}_{0i}(\mu)-\frac{\Tilde{X}}{\varepsilon^2}-\frac{\Tilde{Y}}{\varepsilon} }_{\text{tree}}+\ldots.
\label{sigma_i3}
\ee
Here, the ellipsis denotes terms independent of 
\( \mu \) 
and 
\( \varepsilon \). 
The first term originates from two-loop graphs. It introduces a factor 
\( \mu^{2\varepsilon} \) 
for each integration and features at most a double pole. The second term is linear in the renormalized vertices 
\( \bigg(c^{(2)}_{k j}(\mu)-\frac{b_{kj}^{(2)}}{\varepsilon}\bigg)\) 
from 
$S^{(2)}$ 
and contains the factor
$\alpha_{kj}\mu^{2\varepsilon}/\varepsilon$ 
arising from the one-loop integration.
This term contains a 
$1/\varepsilon^2$ 
pole, in contrast to the usual structure of one-loop graphs in renormalizable theories, where such double poles do not occur. The tree-level contribution comes from 
\( S^{(3)} \) 
and provides the renormalization at this order. Expanding this expression into a series in 
\( \varepsilon \) 
and using the RG equation 
\eqref{rg_1}, 
we obtain
\be
\sigma_i^{(3)} &&  = \
\frac{X}{\varepsilon ^2} + \frac{2 X \log \mu^2 +Y}{\varepsilon } + \left(2 X \log ^2\mu^2 +2 Y \log \mu^2 \right) \nonumber \\
&&- \sum_{k,j} \alpha_{k j} b^{(2)}_{kj}  \left[ \frac{3}{2} \log ^2\mu^2 +\frac{2}{\varepsilon } \log \mu^2 +\frac{1}{\varepsilon ^2}\right]+ c^{(3)}_{0i}(\mu)-\frac{\Tilde{X}}{\varepsilon^2}-\frac{\Tilde{Y}}{\varepsilon}  + \ldots.
\ee
The contributions coming from one and two loops, that contain a product of a pole and a logarithm, 
\( 1/\varepsilon \log \mu^2 \), 
are potential sources of nonlocal divergences and must cancel each other, which leads us to the condition 
$X =  \sum\limits_{k,j}\alpha_{kj} b^{(2)}_{kj}$. 
Moreover, this condition automatically ensures the cancellation of the leading pole 
\( 1/\varepsilon^2 \left( X - \sum\limits_{k,j} \alpha_{kj} b^{(2)}_{kj} \right) \), 
which leads to 
\( \tilde{X} = 0 \). 
Eventually, we find
\be
\sigma^{(3)}_i = \frac{Y-\Tilde{Y}}{\varepsilon}+2Y \log \mu^2 + \frac{1}{2}\sum_{k,j} \alpha_{kj} b^{(2)}_{kj} \log^2 \mu^2 + c_{0 i }^{(3)}(\mu)+\ldots.
\ee
The
\( 1/\varepsilon \) 
pole can be eliminated
by setting
\( Y = \tilde{Y} \). 
In the spirit of 
Sec.~\ref{Sec_renorm}, 
we 
denote this coefficient as 
\(\tilde{Y} \equiv  b_{0i}^{(3)}  \). 
The remaining dependence on 
\( \mu \) 
determines the second RG equation; in differential form we obtain
\be
\mu^2\frac{\partial}{\partial{\mu^2}}c^{(3)}_{0 i}(\mu)= - \sum_{k,s}\alpha_{ks} b^{(2)}_{ks} - 2 b^{(3)}_{0i}, 
\ee
which can be used for calculations at 
the next order. 
%
We can now
determine the coefficients of the first two leading logarithms, which are completely fixed by one-loop calculations:
\be
\sigma^{(2)}_{i} \overset{\rm LL}{=} b^{(2)}_{0i}\log\mu^2, \qquad \sigma^{(3)}_{i} \overset{\rm LL}{=} \frac{1}{2}\sum_{k,j}\alpha_{kj} b^{(2)}_{kj} \log^2\mu^2.
\ee
These equations prescribe the following procedure. At second order, one has to compute the coefficients of the $1/\varepsilon$ poles in the one-loop diagrams shown in Fig.~\ref{Fig_sigma2}. These coefficients directly determine the corresponding 
LL contribution. At third order, it is sufficient to consider the one-loop diagrams shown in Fig.~\ref{Fig_sigma3}, compute their divergent parts for arbitrary $c^{(2)}_{jk}$, and then make the replacements
\be
c^{(2)}_{jk} \to
b^{(2)}_{jk}; \ \ \
\text{and} \ \ \ 
\frac{1} {\varepsilon}\ \to \frac{1}{2} \log\mu^2.
\ee

This algorithm can be generalized to arbitrary order $n$ (see Appendix \ref{App_rec_sheme} for a brief overview of the method). At the $n-$th principal order, the highest pole in the divergent part of the corresponding one-loop diagrams should be calculated for arbitrary $c^{(m)}_{jk}, \, m < n$. After that, the replacements \be c_{kj}^{(m)} \to \beta^{{(m)}}_{kj; 
1-\text{loop}}; \ \ \ \frac{1}{\varepsilon^{n-1}} \to \frac{1}{(n-1)!} \log^{n-1} \mu^2
\ee
should be made. Where $\beta^{{(m)}}_{kj; 
1-\text{loop}}$ is defined in \eqref{beta_definition_RG}.

\begin{figure}[H]
\begin{center}
\includegraphics[width=1\textwidth]{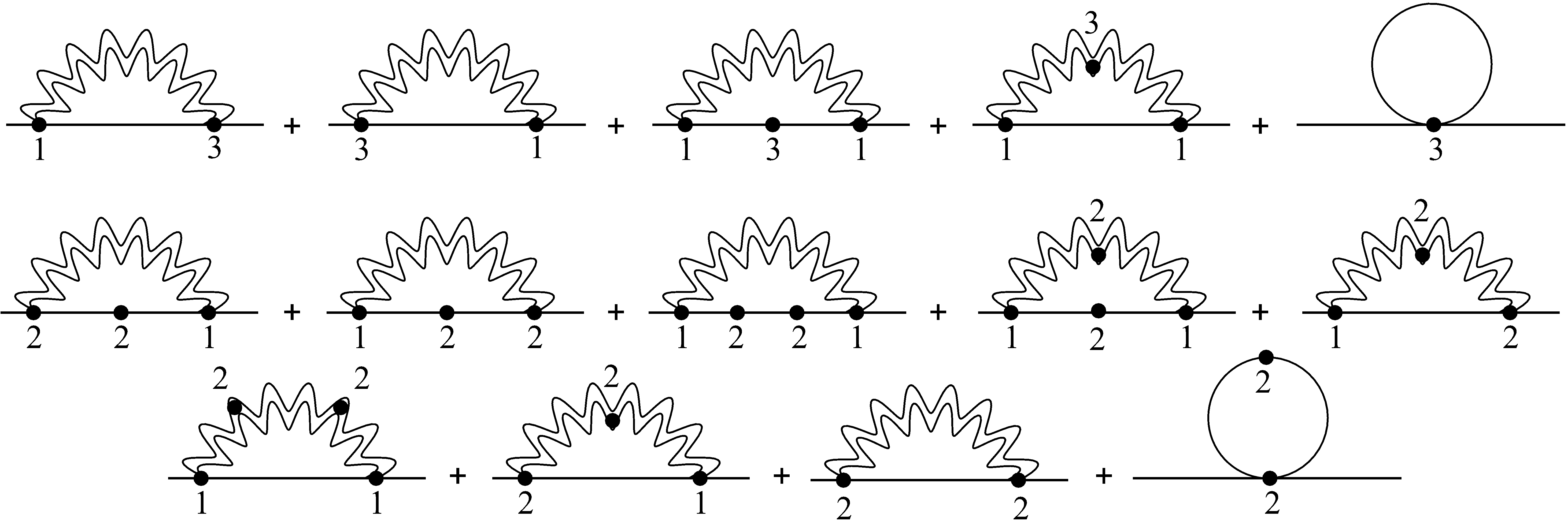} \\
\caption{One-loop diagrams contributing to the running of the scalar mass at the fourth principal order.}
\label{Fig_sigma4}
\end{center}
\end{figure}

The one-loop diagrams entering the calculation at the second and third principal orders were shown in
Figs.~\ref{Fig_sigma2} and~\ref{Fig_sigma3}, respectively. At the fourth principal order, the relevant diagrams are displayed in Fig.~\ref{Fig_sigma4}.

From explicit computation in FORM \cite{Davies:2026cci}, we obtain the LL corrections to $\Sigma$ up to the fourth principal order, corresponding to the three-loop LL contribution:
\be
\Sigma^{(2)}(p^2,m^2) \overset{\rm LL}{=}m^2(p^2-m^2)
L;
\label{Sigma2_LL}
\ee
\be
\Sigma^{(3)}(p^2,m^2) \overset{\rm LL}{=}
\frac{1}{2}\left( \frac{35}{32}m^4 p^2 - \frac{31}{48}m^2 p^4  + \frac{43}{192} p^6 - \frac{41}{32}m^6 \right)
L^2;
\label{Sigma3_LL}
\ee
\be
\Sigma^{(4)}(p^2, m^2)
 \overset{\rm LL}{=}
\frac{1}{6} \bigg(-\frac{3085}{1536} m^8
- \frac{39}{320} p^2 m^6
- \frac{2321}{1440} p^4 m^4
- \frac{5137}{23040} p^6 m^2
+ \frac{12377}{115200} p^8
\bigg) L^3,
\ee
where 
\be
 L \equiv \frac{\kappa^2}{16 \pi^2} \log \frac{\mu^2}{m^2}.
\ee

The physical mass is defined by the pole equation 
\be 
m^2 \,+\,  \Sigma(p^2 = m_{\rm ph}^2, m^2) = m_{\rm ph}^2.
\label{Pole_eq}
\ee
It can be expanded in powers of logarithms
\be 
m_{\rm ph}^2 \overset{\rm LL}{=}  m^2 \left(1 + \sum_{n = 1}^{\infty}m_n L^n \right). \; 
\ee
The recursive solution of the pole equation 
(\ref{Pole_eq})
then yields the LL result for the physical mass up to the three-loop order: 
\be 
m_{\rm ph}^2 \overset{\rm LL}{=} m^2 - \frac{39}{128} m^6 L^2 - \frac{655003}{691200} m^8 L^3 + \mathcal{O}(L^4).
\label{mph_LL}
\ee

The calculation above was performed in the harmonic gauge with $\xi=1$. 
As a nontrivial check of the procedure, we now repeat the calculation in a general harmonic gauge up to the third principal order. 
This is sufficient to verify the gauge-parameter independence of the pole mass at the two-loop LL level. 
Although the off-shell self-energy depends on $\xi$, this dependence must cancel for the solution of
the pole equation since $m_{\rm ph}$ is a physical observable.
The one-loop LL contribution remains
\be
\Sigma^{(2)}(p^2,m^2) \overset{\rm LL}{=}\frac{3-\xi}{2}\,
m^2(p^2-m^2)L ,
\label{Sigma2_LL_xi}
\ee
while at the next order we obtain
\begin{align}
\Sigma^{(3)}(p^2, m^2) \, \,
  \overset{\rm LL}{=} & \, \,
\frac{1}{2}\Big[ m^6 \left(-\frac{57}{32}+ \frac{3}{4}\xi- \frac{1}{4}\xi^2\right)+ p^2 m^4 \left(\frac{29}{32}- \frac{1}{8}\xi+ \frac{5}{16}\xi^2\right)  \nonumber\\
+&\, \, p^4 m^2 \left(-\frac{7}{48}  - \frac{1}{4}\xi - \frac{1}{4}\xi^2\right)
+ p^6 \left(\frac{79}{192} - \frac{3}{8}\xi + \frac{3}{16}\xi^2\right) \Big] L^2.
\label{Sigma3_LL_xi}
\end{align}
Setting $\xi=1$ in Eqs.~\eqref{Sigma2_LL_xi} and~\eqref{Sigma3_LL_xi} reproduces Eqs.~\eqref{Sigma2_LL} and~\eqref{Sigma3_LL}. 
Substitution of Eqs.~\eqref{Sigma2_LL_xi} and~\eqref{Sigma3_LL_xi} into the pole equation~\eqref{Pole_eq} shows that all terms proportional to $\xi$ and $\xi^2$ cancel, which non-trivially verifies our computation of the gauge-invariant pole mass ~\eqref{mph_LL} through the third principal order.

At the fourth principal order the calculation in a general harmonic gauge becomes technically challenging in the present implementation, due to the rapid growth of intermediate tensor expressions, memory consumption, and the running time. For this reason, the $L^3$ term in Eq.~\eqref{mph_LL} is quoted only in the $\xi=1$ gauge. We emphasize that computation of higher terms is also possible without any modification of the procedure, i.e. no new integrals or topologies appear. The only complication is increased volume of index algebra, which makes the computation 
prohibitively time-consuming.

\section{Conclusions and Outlook}
\label{Sec_Conclusions}

In this work, we have developed a framework for the systematic calculation of leading logarithmic (LL) corrections in the effective field theory of gravity. Adapting the renormalization-group techniques previously developed for nonrenormalizable effective theories, in particular Chiral Perturbation Theory, we have shown that the leading logarithms in gravity can be reconstructed recursively using solely one-loop computations and the leading order action $S^{(1)}$.

As a working example, we have considered the scalar two-point function in the effective field theory generated by Einstein-Hilbert action. Using recursive procedure we have obtained the scalar self-energy and the running scalar mass up to the LL of the fourth principal order, corresponding to the three-loop computation. Although the off-shell scalar self-energy depends explicitly on the gauge parameter, we have verified that this dependence cancels in the pole equation, providing a non-trivial check of the consistency of our procedure. As another consistency check of our formulation of gravity, we have analyzed the one-loop graviton polarization operator in a general harmonic gauge and reproduced the standard one-loop quantum correction to the Newtonian potential. 

The present calculation provides the first application of the leading-logarithm recurrence relations to four-dimensional gravity EFT. The recursive approach opens the possibility of studying high-order logarithmic contributions and their resummation in gravitational EFT without brute force evaluation of multiloop gravitational diagrams. A natural next step is to extend this framework to genuinely gravitational observables, such as long-distance gravitational scattering and the quantum corrections to the Newtonian potential, which we plan to consider in subsequent work.

\begin{acknowledgments}
We thank Johan Bijnens,  Fernando Brandt, John Donoghue, Josif Frenkel, Boris Latosh,
Ilya Shapiro, and Dennis  McKeon for useful correspondence.
We are also grateful to Sergey Paston, Alla Semenova and  Vyacheslav Vandeev for enlightening discussions. 

The work of D. Mysliaeva was performed at the Saint Petersburg Leonhard Euler
International Mathematical Institute and supported by the Ministry of Science and Higher
Education of the Russian Federation (agreement no. 075–15–2025–343).

This work was in part supported by Basic Science Research Program through the National Research Foundation of Korea (NRF) funded by the Ministry of Education  RS-2023-00238703 and RS-2018-NR031074.

\end{acknowledgments}

\appendix

\section{Interaction vertices}
\label{append:vertices}

In this Appendix, we collect the interaction vertices generated by the
first-principal-order action 
$S^{(1)}$. 
To this end, we expand
$\sqrt{-g}$ 
and the inverse metric 
$g^{\mu\nu}$ 
in powers of the graviton field 
$h_{\mu\nu}$, 
using
\(
g_{\mu\nu}=\eta_{\mu\nu}+\kappa h_{\mu\nu}.
\) 
The required expansions are
\begin{align}
    \sqrt{-g}&=\sqrt{- \det[(\eta+ \kappa h)_{\nu \mu}]} = \exp\left( \frac{1}{2}\tr[\log\left(\delta+ \kappa h)^{\nu}_{ \mu}\right] \right)= \nonumber \\ &=\exp \left(- \frac{1}{2}\tr \left[ \sum_{k=1}^{\infty}\frac{(-\kappa h_{\nu}^{\rho})^k}{k}\right] \right)  
     = \sum_{n=0}^{\infty} \frac{(-1)^n}{2^n n!}\left(\sum_{k=1}^{\infty}\frac{(-\kappa)^k h_{\nu}^{\rho_1}\ldots h^{\nu}_{\rho_{k-1}}}{k}\right)^{n}, \\
      g^{\mu \nu} &= \eta^{\mu \nu}+\sum_{k=1}^{\infty}(-\kappa)^kh^{\mu}_{\rho_1}\cdot\ldots\cdot h^{\rho_{k-1}}_{\rho_k}\eta^{\rho_k \nu}.
\end{align}
These formulas determine the subsequent expansion of the nonlinear quantities entering the first-principal-order action
\begin{align}
S^{(1)}
=
\int d^4x\,\sqrt{-g}
\left\{
\frac{2}{\kappa^2}R
+g^{\mu\nu}\partial_\mu\phi\,\partial_\nu\phi
-m^2\phi^2
\right\}
+S^{(1)}_{\text{ghost}}.
\end{align}
The interaction vertices are obtained by differentiating 
\(S^{(1)}\) 
with respect to the fields and subsequently setting the fields to zero:
\small
\begin{align}
    &(2\pi)^{4}\delta^{4}(p+p^{\prime}+q)V^{\mu \nu}_{h \phi \phi}= \frac{\delta}{\delta\phi(p)}\frac{\delta}{\delta\phi(p^{\prime})}\frac{\delta}{\delta h_{\mu\nu}(q)}iS^{(1)}, \quad (2\pi)^{4}\delta^{4}(p+p^{\prime}+q)V^{\mu \nu}_{hh \phi \phi}= \frac{\delta}{\delta\phi(p)}\frac{\delta}{\delta\phi(p^{\prime})}\frac{\delta}{\delta h_{\mu\nu}(q)}\frac{\delta}{\delta h_{\rho\sigma}(q')}iS^{(1)}; \nonumber\\
     &(2\pi)^{4}\delta^{4}(p+r+q)V_{hhh}^{\mu \nu \, \lambda \sigma \, \rho \tau}= \frac{\delta}{\delta h_{\lambda \sigma}(q)}\frac{\delta}{\delta h_{\rho\tau}(r)}\frac{\delta}{\delta h_{\mu\nu}(p)}iS^{(1)}, \quad (2\pi)^{4}\delta^{4}(p+p'+q)V^{\mu \, \nu \, \rho \sigma}_{\bar{c} c h}=\frac{\overset{\to}{\delta}}{\delta c_{\nu}(p')}\frac{\overset{\to}{\delta}}{\delta \Bar{c}_{\mu}(p)}\frac{\delta}{\delta h_{\rho\sigma}(q)}iS^{(1)}.
\end{align}
\normalsize
Below we list the explicit expression for the corresponding vertices. 

\subsection{2-scalar-1-graviton and 2-scalar-2-graviton vertices}

\begin{figure}[H]
\begin{center}
\includegraphics[width=0.25\textwidth]{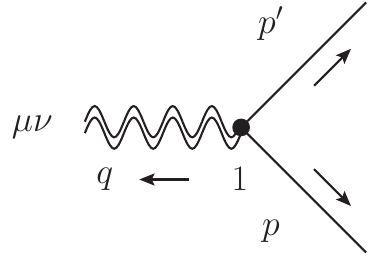} \ \ \ \ \
\includegraphics[width=0.25\textwidth]{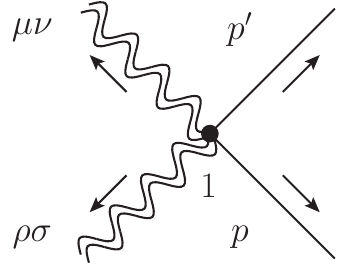}
\end{center}
\caption{Left panel: 2-scalar-1-graviton vertex $V_{h \phi \phi}^{\mu \nu}(p,p')$. Right panel: 2-scalar-2-graviton vertex $V_{hh \phi \phi}^{\mu \nu \, \rho \sigma}(p,p')$. The direction of momenta flow is indicated with arrows. All momenta are set outgoing. }
\label{Fig_V_g_phi}
\end{figure}

Our two-scalar–one-graviton and two-scalar–two-graviton vertices, see Fig.~\ref{Fig_V_g_phi}, coincide with those given in Appendix~A of Ref.~\cite{Bjerrum-Bohr:2002gqz}, up to the convention that all external momenta are defined to be outgoing. The vertices are written as
\be
{ V}_{h \phi \phi}^{\mu \nu}\left(p, p^{\prime} \right)= +
\frac{i \kappa}{2}\left\{p^\mu p^{\prime \nu} +p^\nu p^{\prime  \mu} - \eta^{\mu \nu}\left[\left(p \cdot p^{\prime}\right) + m^2\right]\right\};
\label{Vertex_h_phi_phi}
\ee
and
\be
&&
V_{hh \phi \phi}^{\mu \nu \, \rho \sigma}(p,p')=
i \kappa^2 \Bigg[ 
\left\{ I^{\mu \nu, \,  \alpha \delta} I_{\ \ \ \ \delta}^{\rho, \, \sigma \beta} + \frac{1}{4}\left(\eta^{\mu \nu} I^{\rho \sigma, \, \alpha \beta}+\eta^{\rho \sigma} I^{\mu \nu, \, \alpha \beta}\right)\right\}  
 \left(p_\alpha p_\beta^{\prime}+p_\alpha^{\prime} p_\beta\right)  \nn \\ && + \frac{1}{2}\left(I^{\mu \nu, \, \rho \sigma}-\frac{1}{2} \eta^{\mu \nu} \eta^{\rho \sigma}\right) 
  \left[\left(p \cdot p^{\prime}\right) + m^2\right] \Bigg], 
\label{Vertex_h2_phi_phi}  
\ee
where
\be
I_{\alpha \beta, \, \gamma \delta}=\frac{1}{2}\left(\eta_{\alpha \gamma} \eta_{\beta \delta}+\eta_{\alpha \delta} \eta_{\beta \gamma}\right)
\label{Def_Sym_I}
\ee
is the symmetrized unit tensor. 

\subsection{$3$-graviton vertex}
\begin{figure}[H]
\begin{center}
\includegraphics[width=0.3\textwidth]{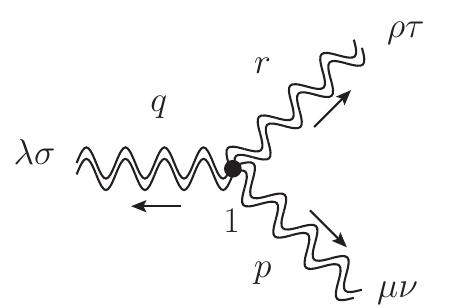}
\end{center}
\caption{$3$-graviton vertex $V_{hhh}^{\mu \nu \, \lambda \sigma \, \rho \tau}(p,q,r)$; all momenta are outgoing. }
\label{Fig_Vh_h_h}
\end{figure}
Here we present the complete form of the $3$-graviton vertex 
$V_{hhh}^{\mu \nu \, \lambda \sigma \, \rho \tau}(p,q,r)$, Fig.~\ref{Fig_Vh_h_h}. 
It differs from the familiar expression derived with help of the background field method in 
Ref.~\cite{Bjerrum-Bohr:2002gqz}, 
and involves $3$ ``quantum'' gravitons maintaining the complete symmetry under simultaneous 
permutations of graviton momenta and of the Lorentz indices associated with gravitons. 
\be
&&
V_{hhh}^{\mu \nu \, \lambda \sigma \, \rho \tau}(p,q,r) = 
\text{Sym} 
\Bigg\{ i \kappa \Big[ p^2 \left(- \frac{1}{32}  \eta^{\lambda  \sigma } \eta^{\mu  \nu } \eta^{\rho  \tau }
+ \frac{1}{8}  \eta^{\lambda  \mu } \eta^{\nu  \sigma } \eta^{\rho  \tau }
+ \frac{1}{16}  \eta^{\lambda  \rho } \eta^{\mu  \nu } \eta^{\sigma  \tau }
- \frac{1}{4}  \eta^{\lambda  \nu } \eta^{\mu  \rho } \eta^{\sigma  \tau } \right) 
+ \bigg(\frac{1}{2} p^{\lambda } p^{\mu } \eta^{\nu  \rho } \eta^{\sigma  \tau }   \nn \\ && 
+ \frac{1}{4} p^{\lambda } p^{\rho } \eta^{\mu  \sigma } \eta^{\nu  \tau }
- \frac{1}{4} p^{\lambda } p^{\sigma } \eta^{\mu  \tau } \eta^{\nu  \rho } 
+ \frac{1}{8} p^{\lambda } p^{\sigma } \eta^{\mu  \nu } \eta^{\rho  \tau }
- \frac{1}{4} p^{\lambda } p^{\mu } \eta^{\nu  \sigma } \eta^{\rho  \tau }
- \frac{1}{4} p^{\lambda } p^{\rho } \eta^{\mu  \nu } \eta^{\sigma  \tau }
+ \frac{1}{32} p^{\mu } p^{\nu } \eta^{\lambda  \sigma } \eta^{\rho  \tau }
- \frac{1}{16} p^{\mu } p^{\nu } \eta^{\lambda  \rho } \eta^{\sigma  \tau } \bigg) \nn \\ && 
+\bigg( \frac{1}{8} p^{\lambda } q^{\mu } \eta^{\nu  \rho } \eta^{\sigma  \tau }
+ \frac{1}{4} p^{\lambda } q^{\rho } \eta^{\mu  \tau } \eta^{\nu  \sigma }
- \frac{1}{16} p^{\lambda } q^{\mu } \eta^{\nu  \sigma } \eta^{\rho  \tau }
- \frac{1}{4} p^{\lambda } q^{\rho } \eta^{\mu  \nu } \eta^{\sigma  \tau }
+ \frac{1}{4} p^{\mu } q^{\lambda } \eta^{\nu  \rho } \eta^{\sigma  \tau }
+ \frac{1}{8} p^{\mu } q^{\nu } \eta^{\lambda  \sigma } \eta^{\rho  \tau } \nn \\ && 
+ \frac{1}{2} p^{\mu } q^{\rho } \eta^{\lambda  \nu } \eta^{\sigma  \tau }
- \frac{1}{4} p^{\mu } q^{\rho } \eta^{\lambda  \sigma } \eta^{\nu  \tau } 
- \frac{1}{8} p^{\mu } q^{\lambda } \eta^{\nu  \sigma } \eta^{\rho  \tau } 
- \frac{1}{4} p^{\mu } q^{\nu } \eta^{\lambda  \rho } \eta^{\sigma  \tau }
- \frac{3}{16} p^{\rho } q^{\tau } \eta^{\lambda  \mu } \eta^{\nu  \sigma }
+ \frac{1}{16} p^{\rho } q^{\tau } \eta^{\lambda  \sigma } \eta^{\mu  \nu } \bigg) \nn \\ && 
+( p \cdot q ) \left(- \frac{1}{32} \eta^{\lambda  \sigma } \eta^{\mu  \nu } \eta^{\rho  \tau } 
+ \frac{3}{32} \eta^{\lambda  \mu } \eta^{\nu  \sigma } \eta^{\rho  \tau }
+ \frac{1}{8} \eta^{\lambda  \rho } \eta^{\mu  \nu } \eta^{\sigma  \tau }
- \frac{3}{8} \eta^{\lambda  \nu } \eta^{\mu  \rho } \eta^{\sigma  \tau } \right)\Big]\Bigg\}_{\substack{
\mu \leftrightarrow \nu \\
\lambda \leftrightarrow \sigma \\
\rho \leftrightarrow \tau
}}+ \left(\text{permutations}\right).
\label{Def_3g_vertex_our}
\ee
In the expression above, the symmetrization over the graviton indices 
$\left(\mu \leftrightarrow \nu, \lambda \leftrightarrow \sigma, \rho \leftrightarrow \tau\right)$ 
is performed without introducing any numerical prefactors. Afterwards, all possible permutations of the external legs 
$(p, \mu, \nu), (q, \lambda, \sigma), (r, \rho, \tau)$ 
are taken into account. The complete expression for this vertex contains 
$495$ 
terms.

\subsection{Antighost-ghost-graviton vertex}
\begin{figure}[H]
\begin{center}
\includegraphics[width=0.30\textwidth]{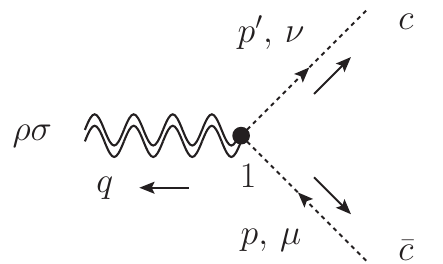}
\end{center}
\caption{Antighost-ghost-graviton vertex; all momenta are outgoing. }
\label{Fig_Vbarc_c_g}
\end{figure} 

The antighost-ghost-graviton vertex, Fig.~\ref{Fig_Vbarc_c_g}, corresponding to fixing of the gauge with help of the term 
(\ref{S1_gf}) 
can be read off the ghost sector action, 
Eq.~(\ref{S_Ghost_sector}):
\be
&&
V^{\mu \, \nu \, \rho \sigma}_{\bar{c} c h}(p,p',q)=
i \kappa \bigg(-\frac{1}{2} {p'}^2 \eta^{\mu \sigma} \eta^{\nu \rho}-\frac{1}{2} {p'}^2 \eta^{\mu \rho} \eta^{\nu \sigma}+\frac{1}{2} {p'}^\mu q^\nu \eta^{\rho \sigma}+\frac{1}{2} {p'}^\rho q^\mu \eta^{\nu \sigma}  \nn \\ && -\frac{1}{2} {p'}^\rho q^\nu \eta^{\mu \sigma}-\frac{1}{2} {p'}^\mu q^\rho \eta^{\nu \sigma}+  
\frac{1}{2} {p'}^\sigma q^\mu \eta^{\nu \rho}-\frac{1}{2} {p'}^\sigma q^\nu \eta^{\mu \rho}-\frac{1}{2} {p'}^\mu q^\sigma \eta^{\nu \rho}-\frac{1}{2} \eta^{\mu \sigma} \eta^{\nu \rho}({p'} \cdot q)-\frac{1}{2} \eta^{\mu \rho} \eta^{\nu \sigma}({p'} \cdot q)  \nn \\ && +\frac{1}{2} q^\mu q^\nu \eta^{\rho \sigma}-\frac{1}{2} q^\nu q^\rho \eta^{\mu \sigma}-\frac{1}{2} q^\nu q^\sigma \eta^{\mu \rho} \bigg).
\ee

\section{Nieuwenhuizen operators}
\label{App_Nieu}

In order to present our result for the one-loop graviton polarization operator in a compact form we employ the set of the so-called 
Nieuwenhuizen operators \cite{VanNieuwenhuizen:1973fi} that provide a complete basis to expand
operators ${\cal O}_{\mu \nu, \, \alpha \beta}(\Delta)$ symmetric in the indices 
$(\mu \nu)$,
$(\alpha \beta)$ and under the interchange of $(\mu \nu)$ with $(\alpha \beta)$.

The Nieuwenhuizen operators are constructed from
the two gauge projectors for gauge models with spin-1:
\be
\theta_{\mu \nu}(\Delta) 
\equiv
\eta_{\mu \nu}-\frac{\Delta_\mu \Delta_\nu}{\Delta^2}, \quad \omega_{\mu \nu}(\Delta) 
\equiv 
\frac{\Delta_\mu \Delta_\nu}{\Delta^2}, \quad \quad \theta_{\mu \nu}(\Delta)+\omega_{\mu \nu}(\Delta)=\eta_{\mu \nu}.
\ee
We use the following definition of the Nieuwenhuizen operators \cite{Accioly:2002tz}:
\be
\begin{aligned}
P_{\mu \nu, \, \alpha \beta}^2(\Delta) & =\frac{1}{2}\left[\theta_{\mu \alpha}(\Delta) \theta_{\nu \beta}(\Delta)+\theta_{\mu \beta}(\Delta)
\theta_{\nu \alpha}(\Delta)\right]-\frac{1}{d-1} \theta_{\mu \nu}(\Delta) \theta_{\alpha \beta}(\Delta), \\
P_{\mu \nu, \, \alpha \beta}^1(\Delta) & =\frac{1}{2}\left[\theta_{\mu \alpha}(\Delta) \omega_{\nu \beta}(\Delta)+\theta_{\mu \beta}(\Delta) \omega_{\nu \alpha}(\Delta)+\theta_{\nu \alpha}(\Delta) \omega_{\mu \beta}(\Delta)+\theta_{\nu \beta}(\Delta) \omega_{\mu \alpha}(\Delta)\right], \\
P_{\mu \nu, \, \alpha \beta}^0(\Delta) & =\frac{1}{d-1} \theta_{\mu \nu}(\Delta) \theta_{\alpha \beta}(\Delta), \\
\bar{P}_{\mu \nu, \, \alpha \beta}^0(\Delta) & =\omega_{\mu \nu}(\Delta) \omega_{\alpha \beta}(\Delta), \\
\bar{\bar{P}}_{\mu \nu, \,\alpha \beta}^0(\Delta) & =\theta_{\mu \nu}(\Delta) \omega_{\alpha \beta}(\Delta)+\omega_{\mu \nu}(\Delta) \theta_{\alpha \beta}(\Delta) .
\end{aligned}
\label{NH_operators}
\ee
Four of these operators form a complete set of projectors and constitute a decomposition of the symmetrized unit tensor (\ref{Def_Sym_I}):
\be
P_{\mu \nu, \, \alpha \beta}^2(\Delta)+P_{\mu \nu, \,  \alpha \beta}^1(\Delta)+P_{\mu \nu,\,  \alpha \beta}^0(\Delta)+\bar{P}_{\mu \nu, \, \alpha \beta}^0(\Delta)=I_{\mu \nu, \, \alpha \beta}.
\ee

The operator $\bar{\bar{P}}$ is orthogonal to $P^1$ and $P^2$. It 
describes the mixing between the two scalar subspaces:

The operator $\bar{\bar{P}}$
is orthogonal to the spin-$2$ and spin-$1$ sectors and acts exclusively in the scalar block. While 
$P^0$
and 
$\bar{P}^0$
 are projectors onto the two independent scalar subspaces, 
$\bar{\bar{P}}$
is off-diagonal in this basis and therefore encodes mixing between them:
\be
P^0 \bar{\bar{P}}^0 P^0=0, \quad \bar{P}^0 \bar{\bar{P}}^0 \bar{P}^0=0.
\ee

We would like to expand a given operator
${\cal O}_{\mu \nu, \, \alpha \beta}(\Delta)$
over the operators (\ref{NH_operators}) as
\be
\mathcal{O}_{\mu \nu, \, \alpha \beta}(\Delta)=A P^2_{\mu \nu, \, \alpha \beta}(\Delta)+B P^1_{\mu \nu, \, \alpha \beta}(\Delta)+C P^0_{\mu \nu, \, \alpha \beta}(\Delta)+D \bar{P}^0_{\mu \nu, \, \alpha \beta}(\Delta)+E \bar{\bar{P}}^0_{\mu \nu, \, \alpha \beta}(\Delta).
\ee
$P^2$ and $P^1$ are real projector operators.
Therefore,
\be
A=  \frac{2}{(d+1)(d-2)} \operatorname{Tr}\left(P^2 \mathcal{O}\right) , \quad B=\frac{1}{d-1}\operatorname{Tr}\left(P^1 \mathcal{O}\right).
\ee
The scalar case is more complicated due to the mixing of the two scalar subspaces. Using some algebra it can be
resolved as
\be
C=\operatorname{Tr}\left(P^0 \mathcal{O}\right), \quad D=\operatorname{Tr}\left(\bar{P}^0 \mathcal{O}\right), \quad E=\frac{1}{d-1}\left(\theta_{\mu \nu}(\Delta) \omega_{\alpha \beta}(\Delta)\right) \mathcal{O}^{\mu \nu, \, \alpha \beta}(\Delta).
\ee 

\section{
Reproducing the One-Loop Quantum Correction to Newton's Law
}
\label{App_one_loop_diags}

In order to verify the consistency of our construction of gravity EFT presented in Sec.~\ref{II},  we calculate the 1-loop quantum corrections to the non-relativistic scattering potential of two masses in the harmonic gauge with $\xi=1$. 

We closely follow the procedure 
of Ref.~\cite{Bjerrum-Bohr:2002gqz}. 
The potential is computed from the $2 \to 2$ scattering amplitude of scalar particles with masses $m_1$ and $m_2$:
\be
\langle k_2 \, k_4 | i T| k_1 k_3 \rangle \equiv(2 \pi)^4 \delta^{(4)}\left(k_1+k_3-k_2-k_4\right) i \mathcal{M}(s,t,u),
\label{M2_to_2}
\ee
where $s=(k_1+k_3)^2$, $t=(k_1-k_2)^2$ and $u=(k_1-k_4)^2$ are the Mandelstam variables. 
Technically, the potential is defined as the Fourier transform of the non-relativistic limit of the amplitude 
\be
V(r)=-\frac{1}{2 m_1} \frac{1}{2 m_2} \int \frac{d^3 \Delta}{(2 \pi)^3} e^{i \vec{\Delta} \cdot \vec{r}} \mathcal{M}(|\vec{\Delta}|),
\ee
where $r \equiv |\vec{r}|$. 
The non-relativistic limit of the $2 \to 2$ amplitude
(\ref{M2_to_2}) is defined as
\be
\mathcal{M}(|\vec{\Delta}|) \equiv  \mathcal{M} \left(s=(m_1+m_2)^2,t=-|\vec{\Delta}|^2 ,u=2(m_1^2+m_2^2)- (m_1+m_2)^2+|\vec{\Delta}|^2 \right).
\ee
Our primary interest lies in the coefficients of the $1/|\vec{\Delta}|$ 
term and of the nonanalytic contribution $\sim \log |\vec{\Delta}|^{2}$ 
in the scattering amplitude, since these pieces generate, respectively, the classical relativistic $1/r^{2}$ 
and the leading quantum 
$1/r^{3}$ 
corrections to the potential. The corresponding coordinate-space behavior follows from the Fourier transforms
\be
\int \frac{d^3 \Delta}{ (2 \pi)^3} e^{i \vec{\Delta} \cdot \vec{r}} \frac{1}{ | \vec{\Delta}|}=   \frac{1}{2 \pi^2 r^2}; \ \ \ 
\int \frac{d^3 \Delta}{ (2 \pi)^3} e^{i \vec{\Delta} \cdot \vec{r}} \log \left( | \vec{\Delta}|^2 \right)= - \frac{1}{2 \pi r^3}.
\label{Non-analytic_conributions}
\ee

\begin{figure}[!ht]
\begin{center}
\includegraphics[height=0.12\textwidth, keepaspectratio]{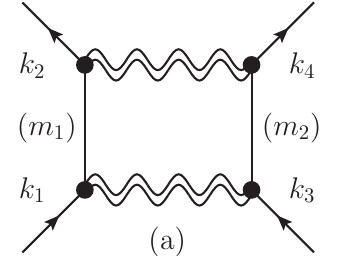} \ \ \ \  
\includegraphics[height=0.12\textwidth, keepaspectratio]{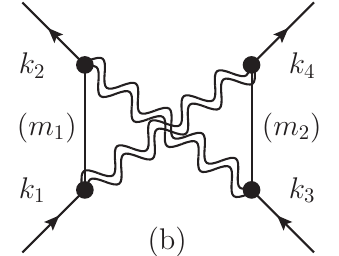}
\end{center}
\caption{The box and crossed box diagrams.}
\label{Fig_Boxes}
\end{figure}

\begin{figure}[!ht]
\begin{center}
\includegraphics[height=0.12\textwidth, keepaspectratio]{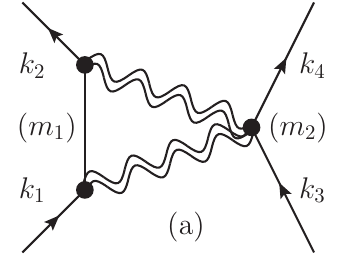} \ \ \ \  
\includegraphics[height=0.12\textwidth, keepaspectratio]{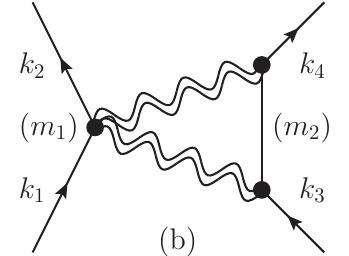}
\end{center}
\caption{The triangle diagrams.}
\label{Fig_triangle}
\end{figure}

\begin{figure}[!ht]
\begin{center}
\includegraphics[height=0.12\textwidth, keepaspectratio]{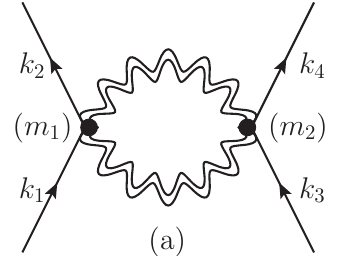} 
\end{center}
\caption{The double seagull diagram.}
\label{Fig_seagull}
\end{figure}

We evaluate the same set of one-loop diagrams as in Ref.~\cite{Bjerrum-Bohr:2002gqz}. The box, triangle and double seagull diagrams depicted, respectively, 
in Figs.~\ref{Fig_Boxes}, \ref{Fig_triangle}, \ref{Fig_seagull},  involve only the  two-scalar–one-graviton and two-scalar–two-graviton vertices (\ref{Vertex_h_phi_phi}), (\ref{Vertex_h2_phi_phi}),
that coincide with those employed in  Ref.~\cite{Bjerrum-Bohr:2002gqz}. Below, for completeness, we quote the corresponding results.

\be
V_{{\rm Fig.}\,\ref{Fig_Boxes}{\rm (a)} +\ref{Fig_Boxes}{\rm (b)} }(r)=- \frac{47}{3} \frac{ G^2  m_1 m_2}{ \pi r^3}.
\label{V_box}
\ee

\be
V_{{\rm Fig.}\,\ref{Fig_triangle}{\rm (a)} +\ref{Fig_triangle}{\rm (b)} }(r)=-4 \frac{G^2 m_1 m_2 (m_1+m_2)}{r^2} +28 \frac{ G^2  m_1 m_2}{ \pi r^3}.
\label{V_triangle}
\ee

\be
V_{{\rm Fig.}\,\ref{Fig_seagull}{\rm (a)} }(r)=- 22 \frac{ G^2  m_1 m_2}{ \pi r^3}.
\ee

Now we consider the vertex correction diagrams, Fig.~\ref{Fig_vertex_corrections}. Since they involve the $ggg$ vertex
(\ref{Def_3g_vertex_our}), their contribution generally differs from that of Ref.~\cite{Bjerrum-Bohr:2002gqz}: 
\begin{figure}[!ht]
\begin{center}
\includegraphics[height=0.12\textwidth, keepaspectratio]{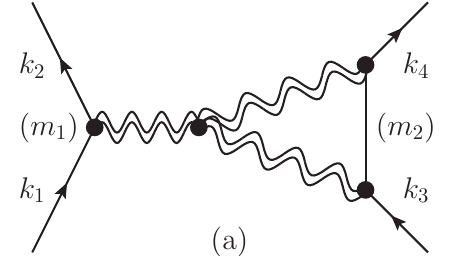} \ \ \ \  
\includegraphics[height=0.12\textwidth, keepaspectratio]
{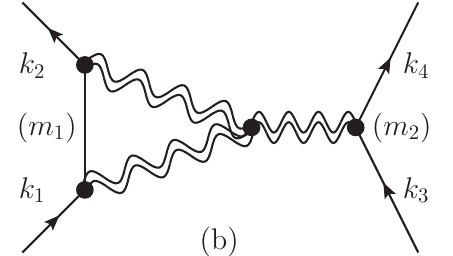} \\ $\,$ \\ 
\includegraphics[height=0.12\textwidth, keepaspectratio]{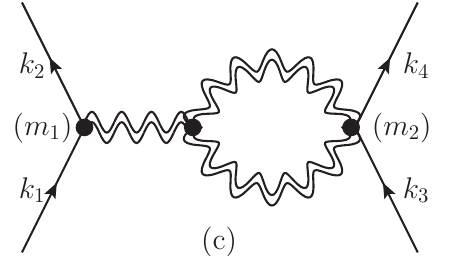} \ \ \ \  
\includegraphics[height=0.12\textwidth, keepaspectratio]
{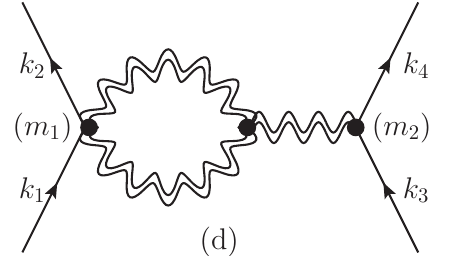} \ \ \ \ 
\end{center}
\caption{The vertex correction diagrams.}
\label{Fig_vertex_corrections}
\end{figure}
\be
V_{{\rm Fig.}\,\ref{Fig_vertex_corrections}{\rm (a)} +\ref{Fig_vertex_corrections}{\rm (b)} }(r)=  \frac{G^2 m_1 m_2 (m_1+m_2)}{r^2} -5 \frac{ G^2  m_1 m_2}{ \pi r^3}.
\label{V_vertex_(a+b)}
\ee
\be
V_{{\rm Fig.}\,\ref{Fig_vertex_corrections}{\rm (c)} +\ref{Fig_vertex_corrections}{\rm (d)} }(r)=  14 \frac{ G^2  m_1 m_2}{ \pi r^3}.
\ee

\begin{figure}[H]
\begin{center}
\includegraphics[height=0.12\textwidth, keepaspectratio]{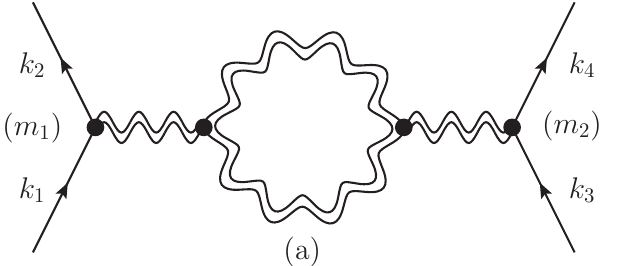} \ \ \ \ 
\includegraphics[height=0.12\textwidth, keepaspectratio]{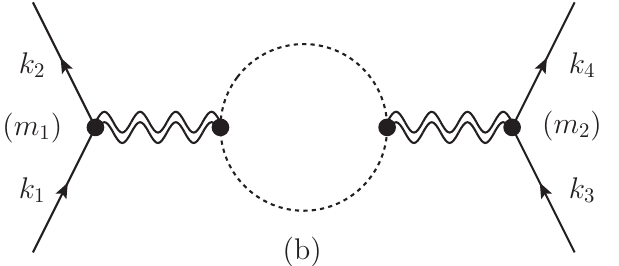}
\end{center}
\caption{The vacuum polarization diagrams. We do not show tadpole graviton diagrams since they vanish in the dimensional regularization. Moreover, we also do not show diagrams with massive loops, as they do not
bring the non-analytic contribution of the type (\ref{Non-analytic_conributions}), and hence do
not contribute into leading corrections to the Newton potential.  }
\label{Fig_Vacuum_pol}
\end{figure}

Finally, by making use of the graviton polarization operator (\ref{GravPol_1PI}), we account for the contribution of
the vacuum polarization diagrams, Fig.~\ref{Fig_Vacuum_pol}.
\be
V_{{\rm Fig.}\,\ref{Fig_Vacuum_pol}{\rm (a)} +\ref{Fig_Vacuum_pol}{\rm (b)} }(r)= -\frac{103}{30} \frac{ G^2  m_1 m_2}{ \pi r^3}.
\label{V_polarization}
\ee



Although the individual contributions from diagrams involving the three-graviton vertex
(\ref{Def_3g_vertex_our}) and the graviton polarization operator (\ref{GravPol_1PI}) differ in their numerical coefficients from those presented in Ref.~\cite{Bjerrum-Bohr:2002gqz}, the sum of all one-loop contributions, (\ref{V_box})-(\ref{V_polarization}), reproduces exactly the same correction to the non-relativistic potential:
\be
V(r)=-\frac{G m_1 m_2}{r}\left[1+3 \frac{G\left(m_1+m_2\right)}{r}+\frac{41}{10 \pi} \frac{G 
}{r^2}\right].
\label{V(r)_Bjerrum_Bohr}
\ee
This agreement provides a nontrivial consistency check of our implementation of the graviton interaction vertices as well as of the ghost sector of the theory 
(\ref{Complete_S_1}).


\section{Leading logarithms from the renormalization group }
\label{App_rec_sheme}
The general framework of RG relations, which underlies  the recursive construction of leading logarithms in effective field theories, was developed in Refs.~\cite{Buchler:2003vw,Colangelo:1995np} and has been extensively used in 
ChPT~\cite{Bissegger:2006ix,Bijnens:2009zi,Bijnens:2010xg,Bijnens:2012hf,Bijnens:2014ila}. In this 
Appendix we summarize the notations and briefly overview the RG relations 
employed in the recursive calculation of leading logarithms in EFTs. 

Following the notation introduced in Eq.~\eqref{Def_Sgrav}, the Lagrangian at the $n$-th principal order can be decomposed over a basis of independent local operators $\{\mathcal{O}_{kj}^{(n)}\}$,
\be
\mathcal{L}^{(n)} = \sum_{k,j} \tilde c^{(n)}_{kj}\, \mathcal{O}^{(n)}_{kj}.
\label{L_decomp_RG}
\ee
Here $k+2$ denotes the number of fields entering the operator, while the index $j$ labels different Lorentz structures. 

In the dimensional regularization, $d=4-2\varepsilon$, the bare coefficients are written as 
\be 
\tilde c^{(n)}_{kj} = \mu^{k\varepsilon} \left[ c^{(n)}_{kj}(\mu) + \sum_{a=1}^{n-1} \frac{A^{(n)}_{kj;a}}{\varepsilon^a} \right]. 
\label{pole_expansion_RG} 
\ee 
Here 
$c^{(n)}_{kj}(\mu)$ 
are renormalized couplings and 
$A^{(n)}_{kj;a}$ 
are pole coefficients. At the $n$-th principal order, diagrams with at most 
$n-1$ 
loops contribute; therefore the pole expansion terminates at 
$a=n-1$. 
The coefficients 
$A^{(n)}_{kj;a}$ 
are polynomial functions of the couplings 
of lower principal orders: 
$ A^{(n)}_{kj;a} \equiv A^{(n)}_{kj;a} \left(c^{(m)}_{li}\right), \, m<n $.
The requirement of independence of the bare Lagrangian 
on the renormalization scale, 
\be
\frac{d\mathcal{L}^{(n)}}{d\log\mu^2}=0,
\ee
relates the pole coefficients to the beta functions. We use the convention 
\be
\beta^{(n)}_{kj} = \frac{d c^{(n)}_{kj}}{d\log\mu^2} + \frac{\varepsilon k}{2}c^{(n)}_{kj}.
\label{beta_definition_RG}
\ee
The general RG equations relate the pole coefficients 
$A^{(n)}_{kj;a}$ 
to the $\beta-$functions of the effective couplings. In the present analysis, we deal only 
with the highest order pole part, 
since it determines the leading logarithms. In this sector, the RG relations take a particularly 
simple recursive form. The coefficient of the highest pole at the $n$-th principal order is given by
\be
A^{(n)}_{kj;n-1}
=
\frac{1}{(n-1)!}
\left(
\sum_{\substack{m<n\\ l,i}}
\beta^{(m)}_{li;{\rm 1\,loop}}
\frac{\delta}{\delta c^{(m)}_{li}}
\right)^{n-2}
\beta^{(n)}_{kj;{\rm 1\,loop}} .
\label{LL_master_formula}
\ee
Here 
$\beta^{(m)}_{li;{\rm 1\,loop}}$ 
denotes the one-loop beta function for the coupling 
$c^{(m)}_{li}$, and the variational derivatives act on the coupling 
dependence of the one-loop pole coefficients. Thus, the leading logarithmic contribution at arbitrary order is 
fixed by the one-loop $\beta-$functions only.

\bibliography{GravityLLog}

\end{document}